\documentclass[conference]{IEEEtran}

\IEEEoverridecommandlockouts

\usepackage{cite}

  \usepackage{graphicx}                
  \DeclareGraphicsExtensions{.eps}     
  
  \usepackage[absolute]{textpos}
\newcommand{\copyrightstatement}{
    \begin{textblock}{15}(0.5,0.3)    
         \noindent
         \centering
         \textblockcolour{white}
         \footnotesize
         \copyright 2019 IEEE. Personal use of this material is permitted. Permission from IEEE must be obtained for all other uses, in any current or future media, including reprinting/republishing this material for advertising or promotional purposes, creating new collective works, for resale or redistribution to servers or lists, or reuse of any copyrighted component of this work in other works
    \end{textblock}
}

\usepackage{amsmath}
\usepackage{psfrag}

\usepackage{array}

\usepackage{url}

\begin{document}

\title{Power Modeling for Virtual Reality \\ Video Playback Applications}

\copyrightstatement

\author{

\IEEEauthorblockN{Christian Herglotz, St\'ephane Coulombe}
\IEEEauthorblockA{ Department of Software Engineering and IT\\
\'Ecole de technologie sup\'erieure (\'ETS)\\
Montr\'eal, Canada
}

\and

\IEEEauthorblockN{Ahmad Vakili}
\IEEEauthorblockA{Summit Tech Multimedia\\
Montr\'eal, Canada\\
www.summit-tech.ca}

\and

\IEEEauthorblockN{Andr\'e Kaup (\textit{Fellow IEEE})}
\IEEEauthorblockA{Multimedia Communications \& \\ Signal Processing, \\
Friedrich-Alexander University \\ Erlangen-N\"urnberg (FAU)\\
Erlangen, Germany}
\vspace{-0.5cm}

}


\maketitle

\begin{abstract}
This paper proposes a method to evaluate and model the power consumption of modern virtual reality playback and streaming applications on smartphones. Due to the high computational complexity of the virtual reality processing toolchain, the corresponding power consumption is very high, which reduces operating times of battery-powered devices. To tackle this problem, we analyze the power consumption in detail by performing power measurements. Furthermore, we construct a model to estimate the true power consumption with a mean error of less than $3.5\%$. The model can be used to save power at critical battery levels by changing the streaming video parameters. Particularly, the results show that the power consumption is significantly reduced by decreasing the input video resolution. 
\end{abstract}

\begin{IEEEkeywords}
360 degree, video coding, smartphone, power
\end{IEEEkeywords}

%
\IEEEpeerreviewmaketitle

\section{Introduction}

Due to the recent advances in virtual reality (VR) techniques as well as the rise in the computing power of mobile devices, most modern smartphones are capable of executing real time VR applications. These applications can implement local playback, online streaming, or interactive gaming scenarios. One of the downsides of running these applications on portable devices is that the battery drains quickly because of high computational requirements. 

This paper presents an initial work on power modeling of video playback in VR platforms using Android-operated smartphones. 
We consider the scenario of a smartphone user watching a video on his device using headsets like Google's cardboard or Samsung's Gear VR. The application performs pure playback such that no user interaction is possible except for basic user commands like play, stop, or pause. We consider two video source scenarios: local playback and online streaming via a Wi-Fi connection. 
Finally, various use cases like traditional 2D videos, $360^\circ$ panorama videos, as well as 3D-$360^\circ$ videos are considered. Testing two different applications ensures that the proposed power model is not restricted to a single application. 

\begin{figure}[t]
\psfrag{C}[c][c]{\small{Control}}
\psfrag{R}[c][c]{\small{File/Network I/O}}
\psfrag{D}[c][c]{\small{Decode}}
\psfrag{S}[c][c]{\small{Sense}}
\psfrag{P}[c][c]{\small{Position}}
\psfrag{B}[c][c]{\small{Render}}
\psfrag{M}[c][c]{\small{Display}}
\includegraphics[width=.48\textwidth]{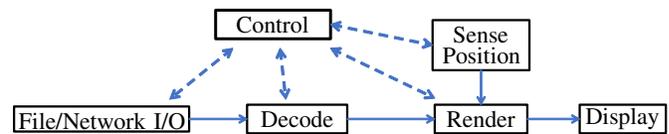}
\vspace{-0.3cm}
\caption{Flowchart of the processing pipeline for VR video playback. }
\label{fig:pipeline}
\vspace{-.3cm}
\end{figure}

Fig.\ \ref{fig:pipeline} shows a high-level diagram of the main processes being performed during VR video playback. 
The control process sends commands to the processing entities and ensures synchronization. This control is usually performed by the central processing unit (CPU).  The video data can be received by either a network interface like Wi-Fi or long-term evolution (LTE), or it can be read from an internal storage like a secure digital (SD) card. The compressed video data is then transferred to the decoder that reconstructs the output image. In this study, we only consider hardware decoding because most modern devices provide such functionality. Furthermore, decoding of high resolution videos (e.g., $4$K) with software is often not feasible in real time. 

Concurrently, a motion sensor like a gyroscope or an accelerometer, which is often available on motion processing units (MPUs), delivers information on the current orientation and motion of the smartphone in real time. The rendering process exploits the motion information and the raw video data to render the output images for the display. These processes are usually performed by the graphics processing unit (GPU). Finally, the output frames are sent to the display. 

Similar studies have already examined various applications like online browsing, video filming, or standard video playback \cite{Carroll13}, where the focus was put on a high variety of applications instead of a detailed analysis on a single application. Furthermore, different hardware modules being active in the VR streaming pipeline such as the 
 network interface \cite{Zou17, Sun14} and the display \cite{Liu16} were investigated. Many studies focused on the video decoding process which can be software-bound \cite{Mallikarachchi17, Herglotz18} or hardware-bound \cite{Li12, Herglotz18a}. A detailed study on the power consumption during online VR video streaming has been performed in \cite{Jiang17}, where the power consumption of a smartphone was analyzed in detail. However, only few test cases with a restricted set of input videos were considered, such that no dependencies on high-level parameters like the frame rate or the resolution of the video were discussed. To the best of our knowledge, this is the first work analyzing the power consumption depending on sequence properties and constructing a respective power model. 

In this work, we focus on the power of data processing such that peripheral devices like the display are neglected. 
 Further work will include such considerations in our power model. However, as the tested device in this paper uses common portable architecture, reported absolute power savings are also valid for portable devices with a display attached. 

 In Section\ \ref{sec:meas}, we introduce the power measurement setup using a smartphone-like test board. Afterwards, in  Section\ \ref{sec:model}, we present the modeling approach that is inspired by power modeling for video streaming. Then, in Section\ \ref{sec:eval}, we validate the proposed models on various video streams and playback settings. The trained models are analyzed and give insights into the power characteristics of the VR video process that are also discussed. Finally, Section\ \ref{sec:concl} concludes this paper.

\section{Power Measurement Setup}
\label{sec:meas}

For the measurement of the power consumption, we chose the Eragon 820 software development kit (SDK) evaluation board \cite{EragonSDK}. It is equipped with a Qualcomm Snapdragon 820 system-on-module (SOM). The main specifications of the SOM are summarized in Table~\ref{tab:820specs}. This system includes a quad-core processor that is used in many modern smartphones of manufacturers. 
\begin{table}[t]
\renewcommand{\arraystretch}{1.3}
\caption{Main specifications of Qualcomm's Snapdragon 820 SOM \cite{EragonSDK}.  }
\vspace{-0.4cm}
\label{tab:820specs}
\begin{center}
\begin{tabular}{l|l}
\hline
Module & Properties \\
\hline
CPU & Qualcomm Kryo quad core ($64$ bit) \\
 & \quad $2$ low power cores (max $1.593$ GHz)\\
  & \quad $2$ high-performance cores (max $2.15$ GHz)\\
GPU & Adreno $530$ (up to $624$ MHz)\\
Memory & $4$ GB of LPDDR $4$ (up to $1866$ MHz) \\
Multimedia & $4$K video decoding at $60$ fps (H.264 or HEVC) \\
\hline
 \end{tabular}
\end{center}
\end{table}

Android is used as the operating system and the impact of simultaneously running subroutines is minimized by uninstalling, disabling, or removing all unnecessary system processes  from the application cache. Furthermore, services such as  Bluetooth, global positioning system (GPS), and Wi-Fi are switched off when not used for measurements. 
To simulate the display output, an external screen is attached to the HDMI output port. 

\begin{figure}[t]
\centering
\psfrag{P}[c][c]{Power Meter \&}
\psfrag{M}[c][c]{Voltage Source}
\psfrag{D}[c][c]{Evaluation Board}
\psfrag{R}[r][r]{Power}
\psfrag{A}[r][r]{Data}
\psfrag{V}[l][l]{USB}
\psfrag{R}[c][b]{Server}
\psfrag{L}[c][c]{Downstream}
\psfrag{U}[l][l]{Upstream}
\psfrag{W}[c][c]{Workstation}
\includegraphics[width=.4\textwidth]{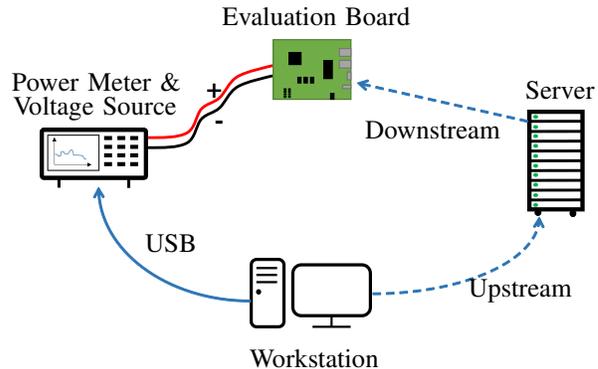}
\vspace{-.2cm}
\caption{Measurement setup with evaluation board, power meter, streaming server, and workstation for data analysis.  }
\label{fig:setup}
\end{figure}

The power itself is measured through the main power supply jack using an external power meter as shown in Fig.\ \ref{fig:setup}. 
The power meter is a Monsoon High Voltage Power Meter (HVPM) \cite{HVPM}. It provides the input voltage of $12$ V and reads the power consumption with a sampling rate of $5\,000$ kHz. To simulate online streaming, the video sequence is streamed via Adobe's RTMP network protocol. The evaluation board receives the live stream through the Wi-Fi interface. 

As tested applications, we choose VaR's VR Player \cite{VaR} and VRTV \cite{VRTV}. The former one is used to measure local playback, the latter one to measure streaming through a Wi-Fi network. Both applications allow various playback options as explained in the following. For each option, a unique identifier is defined. 

First of all, classic 2D video playback is performed in which the video stream is just displayed on the device. Second, a stereo ('st') view is enabled, in which the same image is displayed on the left and the right part of the display, such that a VR headset can be used. In the basic case, head tracking is disabled such that the video is always located exactly in front of the eyes. Next, head tracking is enabled such that the displayed content changes dynamically based on the tracked position ('dyn'), although the screen is at a fixed position in the VR environment. The next option allows to watch 360$^\circ$ sequences, in which the VR user is immersed into the sequence ('360'). 
Finally, 3D VR is tested by using a left and a right view ('3D') to include depth information. 

Further playback options were tested in the dynamic case. Regarding head tracking, in the standard setting, the sensor is chosen automatically. Additional measurements were performed for the explicit use of the gyroscope ('gyro'), the accelerometer ('accel'), and the magnetometer ('magn'). It is worth mentioning that head tracking only worked correctly in standard and in gyroscope mode. 

Furthermore, lens distortion, zooming, and physical head movement using a turntable were tested. 
As the power measurements indicate that all these options have a very small influence on the power (the mean power for all test sequences differed by less than $1\%$ in comparison to the default setting), they are not used and tested in the proposed modeling approaches. 
Table\ \ref{tab:playbackSettings} summarizes the eight playback option sets that were measured.

\begin{table}[t]
\renewcommand{\arraystretch}{1.3}
\caption{List with measured playback settings $\mathrm{I}$ to $\mathrm{VIII}$. '$\mathrm{A1}$' corresponds to VaR's VR Player and '$\mathrm{Ab}$' to both applications. '$\mathrm{x}$' and '$\mathrm{-}$' indicate that the options are switched on and off, respectively. The indicator '$\mathrm{a}$' for the rows '$\mathrm{gyro}$', '$\mathrm{accel}$' and '$\mathrm{magn}$' indicate that the head-tracking sensor is chosen automatically by the application. }
\vspace{-0.4cm}
\label{tab:playbackSettings}
\begin{center}
\begin{tabular}{l|c|c|c|c|c|c|c|c}
\hline
 & $\mathrm{I}$ & $\mathrm{II}$ & $\mathrm{III}$& $\mathrm{IV}$& $\mathrm{V}$ & $\mathrm{VI}$& $\mathrm{VII}$& $\mathrm{VIII}$\\
\hline
Apps & A1 & A1& Ab& Ab& Ab& A1& A1& A1\\
st & -&  x &x&x&x&x&x&x\\
dyn & -&      -&x&x&x&x&x&x\\
$360$ & - &   -&-&x&x&x&x&x\\
$3$D & -&     -&-&-&x&x&x&x\\
gyro & -&     -&a&a&a&x&-&-\\
accel & -&    -&a&a&a&-&x&-\\
magn & -&  -&a&a&a&-&-&x\\
\hline

 \end{tabular}
\end{center}
\end{table}

\begin{figure}[t]
\centering
\psfrag{000}[c][c]{$0$}
\psfrag{001}[c][c]{$2$}
\psfrag{002}[c][c]{$4$}
\psfrag{003}[c][c]{$6$}
\psfrag{004}[c][c]{$8$}
\psfrag{005}[c][c]{$10$}
\psfrag{006}[c][c]{$12$}
\psfrag{007}[c][c]{$0$}
\psfrag{008}[c][c]{$2$}
\psfrag{009}[c][c]{$4$}
\psfrag{010}[c][c]{$6$}
\psfrag{011}[c][c]{$8$}
\psfrag{012}[t][t]{Time [s]}
\psfrag{013}[b][t]{Power [W]}
\includegraphics[width=.48\textwidth]{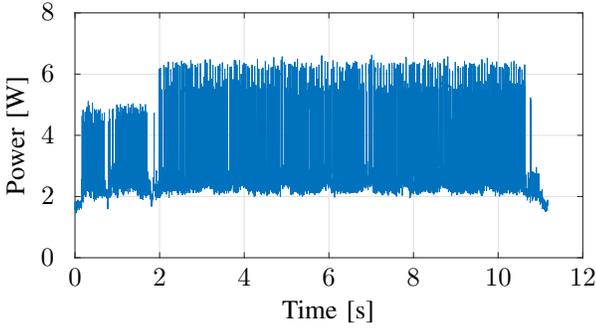}
\vspace{-.3cm}
\caption{Power consumption of the VR playback process including head tracking using VaR's VR Player (sequence AerialCity at $3840\times 1920$ pixels, $30$ fps, crf $28$, cf. Table\ \ref{tab:seqs}). }
\label{fig:power}
\vspace{-.2cm}
\end{figure}

A showcase power measurement for watching $10$ s of video using VaR's VR Player with head tracking enabled and a $360^\circ$ video format is shown in Fig.\ \ref{fig:power}. 	
The x-axis shows the time in seconds (s) and the y-axis the measured power values in watts (W). At the beginning, the app is launched and waits for the play command. The playback starts at approximately $0.1$~s. A short initialization phase follows until $2$~s, during which the maximum power is less than $6$~W. Finally, after $2$~s, the power reaches a static behavior with minimum and maximum powers of $2$~W and more than $6$~W, respectively. The playback ends at approximately $11.5$~s. 

Further tests showed that for longer playback cases, the power characteristics do not change. 
 To make sure that the mean power is averaged over the static playback process, only $7$~s of the complete process, which begins after the initialization phase and ends before the stop, are considered.

To have more realistic power results, a constant offset power is subtracted from all power measurements. This constant offset corresponds to the power the evaluation board consumes in idle mode. In this case, idle mode is considered to be the case when no user application is running, i.e., the display shows the home screen. The advantage is that the differential power values are comparable to the power consumption of real smartphones that are not affected by overhead hardware modules like Ethernet connectors or additional USB ports.

\section{Power Modeling}
\label{sec:model}

We assume that each power component adds to the complete power linearly and 
 write the complete power $\hat P_\mathrm{VR}$ as 
\begin{equation}
\hat P_\mathrm{VR} = \hat P_\mathrm{cont} + \hat P_\mathrm{rec} + \hat P_\mathrm{dec} + \hat P_\mathrm{sens} + \hat P_\mathrm{rend} + \hat P_\mathrm{disp}, 
\label{eq:VRgeneral}
\end{equation}
where the circumflex indicates that the power is estimated. $\hat P_\mathrm{cont}$ is the power needed for controlling the process, $\hat P_\mathrm{rec}$ is the receiver power, $\hat P_\mathrm{dec}$ the decoding power, $\hat P_\mathrm{sens}$ the sensing power, $\hat P_\mathrm{rend}$ the rendering power, and $\hat P_\mathrm{disp}$ the power of the display. This approach corresponds to the main processing steps identified in Fig.\ \ref{fig:pipeline}. 

For the controlling power, no explicit model is known from the literature. For simplicity, we assume a constant power 
\begin{equation}
\hat P_\mathrm{cont} = p_\mathrm{cont,0}. 
\label{eq:control}
\end{equation}
 In terms of the receiver power $\hat P_\mathrm{rec}$, when fixed network settings are used, Sun et al. \cite{Sun14} found that a constant offset and a linear term depending on the bitrate $b$ are the main factors describing the receiver power in Wi-Fi networks. A similar observation was made by Huang et al. \cite{Huang12} for 4G LTE networks. 
 Hence, we model the receiver power as  
\begin{equation}
\hat P_\mathrm{rec} = p_{\mathrm{rec},b}\cdot b + p_{\mathrm{rec},0}, 
\end{equation}
where both parameters can have different values for different networks. 

For the decoding power, we consider a model that originally targets energy estimation  \cite{Herglotz18a}. The model considers the variables frame rate $f$, resolution $S$, and bitrate $b$. We rewrite the model to enable power estimation and obtain 
\begin{equation}
\hat P_\mathrm{dec} = p_\mathrm{dec,0} + p_{\mathrm{dec},f}\cdot f + p_{\mathrm{dec},S}\cdot S + p_{\mathrm{dec},b}\cdot b. 
\label{eq:Me}
\end{equation}
In this model, the parameter $p_\mathrm{dec,0}$ is a constant offset power and the parameters $p_{\mathrm{dec},f}$, $p_{\mathrm{dec},S}$, and $p_{\mathrm{dec},b}$ describe linear contributions of $f$, $S$, and $b$. The resolution $S$ is the product of the pixel width and the pixel height. 

For rendering, we consider the power consumption of the GPU. A corresponding power analysis was performed by Chen et al. \cite{Chen13}. However, a suitable model was not proposed. 
In this work, we propose to also use the frame rate and the resolution of the sequence for power modeling because these two metrics are directly related to the computational complexity of the rendering process. 
 As both parameters are already included in the decoder power models, we assume that no additional modeling parameter is required for $\hat P_\mathrm{rend}$. As a consequence, the decoder parameters cover both decoding and rendering powers. 

Next, the options of the applications discussed in Section\ \ref{sec:meas} are considered ('dyn', '360', etc.). The settings affect both the rendering and the sensing process. Our measurements indicate that it is sufficient to model these settings using constants as 
\begin{align}
\hat P_\mathrm{sens} =  p_\mathrm{gyro}\cdot F_\mathrm{gyro}
 + p_\mathrm{accel}\cdot F_\mathrm{accel} + p_\mathrm{magn}\cdot F_\mathrm{magn} \label{eq:VRmodel_sens}
\end{align}
and
\begin{align}
\hat P_\mathrm{rend} = p_\mathrm{st}\cdot F_\mathrm{st}  + p_\mathrm{dyn}\cdot F_\mathrm{dyn} 
  + p_\mathrm{360}\cdot F_\mathrm{360} +  p_\mathrm{3D}\cdot F_\mathrm{3D}.\label{eq:VRmodel_rend} 
\end{align}
The parameters $F_{(\cdot)}$ represent flags showing with the values $1$ and $0$ whether the setting is activated. For example, $F_\mathrm{st}=1$ means that two views for the use with a VR headset are rendered and $F_\mathrm{st}=0$ means that only the classic single view (like in regular video streaming) is shown on the display. 

Finally, as the display is not part of the measurements, we neglect its power consumption ($\hat P_\mathrm{disp}=0$). 
We gather the information from \eqref{eq:VRgeneral} to \eqref{eq:VRmodel_rend} and combine the parameters that are linearly redundant. These are the constant offsets 
\begin{equation}
p_0 =   p_{\mathrm{cont,}0} + p_{\mathrm{rec,}0} + p_{\mathrm{dec,}0} 
\end{equation}
and the bitrate dependent parameters
\begin{equation}
p_b = p_{\mathrm{rec,}b} + p_{\mathrm{dec,}b}. 
\end{equation}
The resulting model reads
\begin{align}
\hat P_\mathrm{VR,a} = & p_{0} + p_{b}\cdot b +  p_{\mathrm{dec}, f} \cdot f + p_{\mathrm{dec},S} \cdot S \label{eq:VRmodel2}\\ 
 &+ p_\mathrm{st}\cdot F_\mathrm{st}  + p_\mathrm{dyn}\cdot F_\mathrm{dyn} 
  + p_\mathrm{360}\cdot F_\mathrm{360} +  p_\mathrm{3D}\cdot F_\mathrm{3D}\,  
\notag\\
&+p_\mathrm{gyro}\cdot F_\mathrm{gyro}
 + p_\mathrm{accel}\cdot F_\mathrm{accel} + p_\mathrm{magn}\cdot F_\mathrm{magn}. \notag 
 \end{align}
It includes $10$ variables and $K=11$ parameters and in the following, is referred to as advanced model.

Evaluations showed that within the limits of the content we tested, several variables only contribute marginally to the modeling accuracy such that they can be dropped. Hence, we construct a simplified model with the following method. We use \eqref{eq:VRmodel2} as a baseline, drop one by one each of the input variables, and discard those that cause an estimation error increase by less than $0.5\%$. The resulting model reads 
\begin{equation}
\hat P_\mathrm{VR,s} = p_{0} + b\cdot p_{b} +  p_{\mathrm{dec},S} \cdot S   + p_\mathrm{360}\cdot F_\mathrm{360},
\label{eq:VRmodel4}
\end{equation}
which includes $3$ variables and $K=4$ parameters. In the following, this model is called simplified model. It comprises the parameters with the highest impact on the power consumption.  

\section{Experimental Results}
\label{sec:eval}

The evaluation is split into four parts. First, we present the set of tested input video sequences in Section\ \ref{secsec:testseqs}. Afterwards, we discuss the evaluation method which makes use of the mean estimation error in Section\ \ref{secsec:evalMethod}. Third, we evaluate the models from Section\ \ref{sec:model} in Section\ \ref{secsec:errors} and finally, the modeling results are interpreted in Section\ \ref{secsec:interp}. 

\subsection{Test Sequences}
\label{secsec:testseqs}

We measure the power consumption of the two applications with a high number of input sequences with different properties. These sequence-related properties comprise the content of the sequence, the frame rate, the resolution, and the format projection. As for the format projections, we choose classic rectilinear (recti) videos and equirectangular (equi) videos representing 360$^\circ$ content. Additionally, the top-bottom equirectangular format that allows 3D-360$^\circ$ playback is used. The rectilinear and equirectangular sequences are taken from the HEVC common test conditions \cite{Bossen13} and from VVC documents \cite{JVET_K1012}, respectively. A personal archive was used for the 3D sequences, which were all recorded from a fixed camera position. The sequences and their main properties are listed in Table\ \ref{tab:seqs}. 

\begin{table}[t]
\renewcommand{\arraystretch}{1.3}
\caption{Test sequences used for power measurements. All sequences have a duration of $10$ s.  }
\label{tab:seqs}
\vspace{-0.5cm}
\begin{center}
\begin{tabular}{l|c|c|c|c}
\hline
Name & $S$ & $f$  & Projection & 3D \\
\hline
BQSquare & $416\times 240$ & $60$ & recti & no \\
BlowingBubbles & $416\times 240$ & $50$ & recti & no \\
BasketballPass& $416\times 240$ & $50$ & recti	 & no \\
RaceHorses& $416\times 240$ & $30$ & recti & no \\
\hline
BQMall& $832\times 480$ & $60$ & recti & no \\
BasketballDrill & $832\times 480$ & $50$ & recti & no \\
PartyScene& $832\times 480$ & $50$ & recti & no \\
Flowervase& $832\times 480$ & $30$ & recti & no \\
\hline
FourPeople& $1280\times 720$ & $60$ & recti & no \\
Johnny & $1280\times 720$ & $60$ & recti& no \\
SlideEditing& $1280\times 720$ & $30$ & recti & no \\
SlideShow& $1280\times 720$ & $20$ & recti& no \\
\hline
BQTerrace& $1920\times 1080$ & $60$ & recti & no \\
BasketballDrive& $1920\times 1080$ & $50$ & recti & no \\
Cactus& $1920\times 1080$ & $50$ & recti & no \\
Kimono& $1920\times 1080$ & $24$ & recti & no \\
\hline
AerialCity & $3840\times 1920$ & $30$ & equi & no \\
DrivingInCity & $3840\times 1920$ & $30$ & equi & no \\
DrivingInCountry & $3840\times 1920$ & $30$ & equi & no \\
PoleVault & $3840\times 1920$ & $30$ & equi & no \\
\hline
Cars02 & $3840\times 2160$ & $30$ & equi & yes \\
Kitchen2 & $3840\times 2160$ & $30$ & equi & yes \\
Skatedance & $4096\times 2048$ & $30$ & equi & yes \\
Wall6 & $3840\times 1920$ & $30$ & equi & yes \\
\hline
 \end{tabular}
\end{center}
\end{table}

The 'Cars02' sequence was recorded from a sidewalk showing several cars passing by. The 'Kitchen2' sequence was recorded inside a kitchen showing several people working. The 'Skatedance' sequence was recorded in the center of an ice rink showing girls practicing figure skating. Finally, the 'Wall6' sequence was recorded in a climbing hall. 

The sequences are encoded with the x265 and the x264 encoders \cite{x265, x264}. These encoders were selected to compare the behavior of the two codecs H.264 and HEVC. The standard encoder settings are used in general; however, for each sequence, four instances are coded with the constant rate factors (crf) $18$, $23$, $28$, and $33$ to take different bitrates into account. In total, we use $192$ sequences for our measurements, i.e., $96$ for each codec.

Bit streams with a bitrate higher than 12 Mbps are discarded in measuring the VRTV online streaming case because a stable stream could not be established through the Wi-Fi connection. 
The resulting set includes $78$ bit streams with at least two bit streams for each sequence from Table\ \ref{tab:seqs}. Furthermore, all sequences are tested for all configurations although some configurations would not constitute a reasonable use case, e.g., using a rectilinear video in a 360$^\circ$ environment. This enables to strictly separate the 360$^\circ$-rendering power from the sequence specific decoding power. 

\subsection{Evaluation Method}
\label{secsec:evalMethod}

For model evaluation, we transform the advanced model 
and the simplified model 
 into a vector notation and obtain 
\begin{equation}
 \boldsymbol{\hat P}_{\mathrm{VR}, i} = \boldsymbol{A}_i \cdot \boldsymbol{p}_i, 
\end{equation}
where $i$ indicates the model index ($i\in \{\mathrm{a},\mathrm{s}\}$), $\boldsymbol{p}_i$ is a vector containing all $K$ parameters $p_{(.)}$, $\boldsymbol{A}_i$ is a matrix containing the values of all variables for all $N$ measurements, and $\boldsymbol{\hat P}_{\mathrm{VR}, i}$ a vector containing all $N$ estimated powers when using model $i$. 

To separate the training data from the evaluation data, we perform cross-validation. For each validation iteration, we use the measurements corresponding to one source sequence for validation and all the remaining measurements for training. 
For model training in each iteration of the cross-validation, the vector of model parameters is determined to minimize the sum of the squared estimation errors as 
\begin{equation}
\min_{\boldsymbol{p}_i\in {\rm I\!R}^K} \left[ e_i = \sum_{j=1}^n \left(\boldsymbol{\hat P}_{\mathrm{VR,}i}(j) - \boldsymbol P(j)\right)^2 \right], 
\label{eq:minimize}
\end{equation}
where $j$ is the measurement index, $n$ the cardinality of the training set, $\boldsymbol{\hat P}_{\mathrm{VR,}i}(j)$ the estimated power for the $j$-th measurement using model $i$, and $\boldsymbol P(j)$ the measured power of the $j$-th measurement. The training algorithm is a trust-region reflective algorithm proposed by Coleman et al. \cite{Coleman96}. 

Finally, the models are evaluated using the mean relative estimation error and the maximum relative estimation error from validation as 
\begin{equation}
\bar \varepsilon_i = \frac{1}{N}\sum_{j=1}^N \left| \frac{\boldsymbol{\hat P}_{\mathrm{VR,}i}(j) - \boldsymbol P(j)}{\boldsymbol P(j)}\right|
\end{equation}\
and
\begin{equation}
\varepsilon_{i,\mathrm{max}} = \max\limits_{1\le j\le N} \left\{\left|\frac{\boldsymbol{\hat P}_{\mathrm{VR,}i}(j) - \boldsymbol P(j)}{\boldsymbol P(j)}\right| \right\}. 
\end{equation}

\subsection{Estimation Errors}
\label{secsec:errors}
Table\ \ref{tab:errors} summarizes the estimation errors for VaR's VR player and VRTV separately. 
\begin{table}[t]
\renewcommand{\arraystretch}{1.3}
\caption{Estimation errors for the models introduced in Section\ \ref{sec:model}. The number of trained parameters of the model is indicated in the third column. The fourth column indicates the application that was targeted during cross-validation. }
\label{tab:errors}
\vspace{-0.3cm}
\begin{center}
\begin{tabular}{l|l|l|c|r|r}
\hline
Model & Eq.& \# Parameters & App & $\bar \varepsilon$  & $\varepsilon_\mathrm{max}$ \\
\hline
$\hat P_\mathrm{VR,a}$ & \eqref{eq:VRmodel2} & $10$ & VaR & $2.32\%$ & $13.3\%$ \\
$\hat P_\mathrm{VR,s}$ & \eqref{eq:VRmodel4} & $4$  & VaR & $2.25\%$ & $13.7\%$ \\
\hline
$\hat P_\mathrm{VR,a}$ & \eqref{eq:VRmodel2} & $10$ & VRTV & $3.28\%$ & $14.0\%$ \\
$\hat P_\mathrm{VR,s}$ & \eqref{eq:VRmodel4} & $4$  & VRTV & $3.47\%$ & $13.5\%$ \\
\hline
 \end{tabular}
\end{center}
\end{table}
The table shows that the mean estimation error is below $3.5\%$ for all tested cases.  
 The maximum estimation error is below $15\%$. 

 The estimation errors for the VRTV application are generally larger (over $3\%$), which can be explained by Wi-Fi streaming which is less predictable than local memory reading. 
 The observation that estimation errors of the second model only differ slightly in comparison with the first model confirms that the dropped parameters can be neglected for accurate modeling. 

\subsection{Interpretation}
\label{secsec:interp}

%

\begin{figure}
\centering
\psfrag{000}[c][c]{$0\%$}
\psfrag{001}[c][c]{$20\%$}
\psfrag{002}[c][c]{$40\%$}
\psfrag{003}[c][c]{$60\%$}
\psfrag{004}[c][c]{$80\%$}
\psfrag{005}[c][c]{$100\%$}
\psfrag{006}[r][r]{\small $p_\mathrm{magn}\cdot F_\mathrm{magn}$}
\psfrag{007}[r][r]{\small $p_\mathrm{accel}\cdot F_\mathrm{accel}$}
\psfrag{008}[r][r]{\small $p_\mathrm{gyro}\cdot F_\mathrm{gyro}$}
\psfrag{009}[r][r]{\small $ p_\mathrm{3D}\cdot F_\mathrm{3D}$}
\psfrag{010}[r][r]{\small $ p_\mathrm{360}\cdot F_\mathrm{360}$}
\psfrag{011}[r][r]{\small $p_\mathrm{dyn}\cdot F_\mathrm{dyn} $}
\psfrag{012}[r][r]{\small $p_\mathrm{st}\cdot F_\mathrm{st}  $}
\psfrag{013}[r][r]{\small $p_{b}\cdot b$}
\psfrag{014}[r][r]{\small $ p_{\mathrm{dec},S} \cdot S$}
\psfrag{015}[r][r]{\small $p_{\mathrm{dec,}f}\cdot f$}
\psfrag{016}[r][r]{\small $p_0$}
\psfrag{017}[t][b]{Maximum Contribution $C_{V,\max}$ }
\includegraphics[width=.48\textwidth]{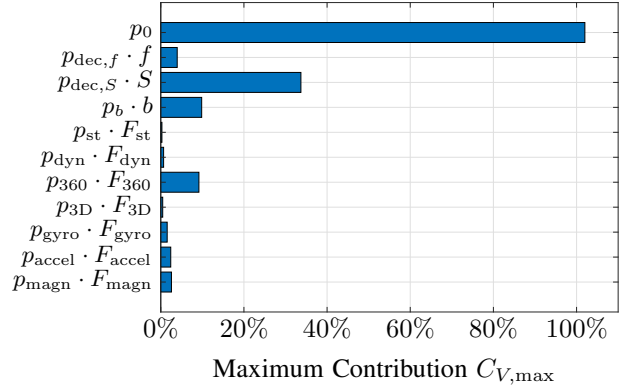}
\caption{Relative maximum contributions $C_{V,\max}$ of parameter-variable products to the complete power.  }
\label{fig:paramImportance}
\vspace{-0.3cm}
\end{figure}

To assess the contribution of each modeling parameter in detail, we calculate their maximum contribution on all measured powers $\boldsymbol P(j)$ using the advanced model. It is obtained by
\begin{equation}
C_{k,\max} [\% ] =  \max\limits_{1\le j \le N} \ \left|\frac{\boldsymbol{A}(j,k)\cdot \boldsymbol{p}(k)}{\boldsymbol{P}(j)}\right| \cdot 100\%,
\label{eq:modelContrib}
\end{equation}
where $k$ is the index of the variable used for modeling (e.g., the frame rate $f$, the resolution $S$, or a flag $F_{(.)}$). $\boldsymbol p(k)$ is the corresponding model parameter as returned by training. The resulting values for all parameters in the case of VaR's VR Player are shown in Fig.\ \ref{fig:paramImportance}. 
We can see that within the limits of the content we tested, the offset power is most important as in the maximum case, the estimated contribution is even higher than the complete measured power ($>100\%$).  The corresponding measurement was made for a low-resolution sequence with a small bitrate and a small frame rate. 

This counter-intuitive observation is obtained because the measurement method does not allow to measure the offset separately from other variables. If VR playback is running, other variables like frame rate and resolution must be non-zero, too. As a consequence, the offset value, which fits the measurement best in terms of the error criterion \eqref{eq:minimize}, can be larger than the smallest measured power. 

Further important variables are the resolution, the bitrate, and the flag indicating $360^\circ$ processing, which all show a maximum contribution of more than $9\%$. A striking observation is that the influence of the frame rate is rather small. From observations in  decoder modeling \cite{Herglotz18a, Li12}, a higher influence would have been expected. 
Presumably, the reason is that for VR applications, the output frame rate is always set to the highest level to be able to follow head rotation as quickly as possible. Hence, differences in the frame rate of the input sequence only affect the decoding process. 

\begin{figure}
\centering
\psfrag{000000000000000000000000}[l][l]{Measured power $P$}
\psfrag{001}[l][l]{$p_0$}
\psfrag{002}[l][l]{$p_S\cdot S$}
\psfrag{003}[l][l]{$p_b\cdot b$}
\psfrag{004}[l][l]{$p_{360}\cdot F_{360}$}
\psfrag{005}[c][c]{$0$}
\psfrag{006}[c][c]{$0.5$}
\psfrag{007}[c][c]{$1$}
\psfrag{008}[c][c]{$1.5$}
\psfrag{010}[r][r]{PartyScene}
\psfrag{009}[r][r]{ crf $28$, $360^\circ$}
\psfrag{012}[r][r]{DrivingInCountry}
\psfrag{011}[r][r]{ crf $28$, $360^\circ$}
\psfrag{014}[r][r]{PartyScene}
\psfrag{013}[r][r]{ crf $28$, no $360^\circ$}
\psfrag{016}[r][r]{DrivingInCountry}
\psfrag{015}[r][r]{ crf $28$, no $360^\circ$}
\psfrag{017}[t][b]{Power [W]}
\includegraphics[width=.47\textwidth]{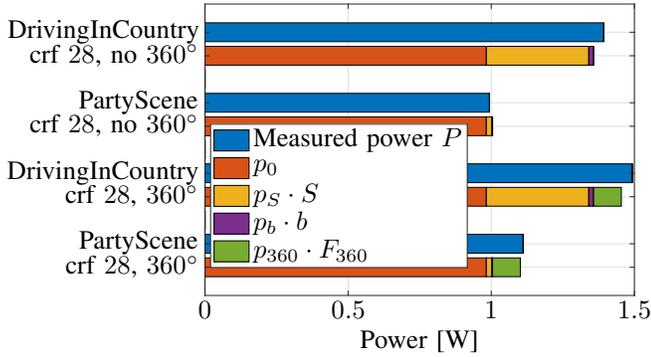}
\caption{Measured and estimated power consumption for four test cases in  VaR's VR Player. The blue bars correspond to the measured power, the stacked bars below the blue bars represent the corresponding power estimates by model $4$, split up into the summands. Head tracking is switched on with standard sensors, the codec is HEVC. }
\label{fig:powerbar}
\end{figure}

To visualize the power consumption for two representative test cases, we plot the measured power and the modeled power distribution in Fig.\ \ref{fig:powerbar}. 
The measured power is depicted by the dark blue bars. The stacked bars below show the power distribution as returned by model $4$. Assuming that modeling is accurate, most of the power is attributed to the offset (almost $1$~W). For the $4$K sequence, the resolution still accounts for approximately $0.4$~W. The bitrate has a rather small influence (due to the high constant rate factor) and the projection format conversion requires approximately $100$~mW. 

These results indicate that a significant amount of power can be saved by switching to a low resolution for the input video. In case a smartphone recognizes that its battery is low, it could switch from a $4$K to a $2$K sequence, which would result in estimated power savings of 
\begin{equation}
\Delta p= \left( 3840\cdot 1920 - 1920\cdot 1080\right) \cdot p_S. 
\label{eq:powerSavings}
\end{equation}
Taking the example of the DrivingInCountry sequence in the $360^\circ$ mode (Fig.\ \ref{fig:powerbar}), this would save $0.266$ W, which accounts for $17.2\%$ of the total power consumption. 

To verify this claim, we encoded the corresponding sequence with the lower resolution of $2$K and measured the corresponding power. A true power reduction of $18.3\%$ could be observed. As similar observations could be made for other sequences, we can conclude that reducing the resolution is a valid and preferred way to reduce the power consumption. 

\section{Conclusion}
\label{sec:concl}
This paper demonstrates that the power needed for live VR applications on mobile devices mainly depends on few parameters: the input video resolution, the bitrate, and the projection format conversion from ERP to rectilinear. This result is verified for two different VR applications. Two models were tested that both reach mean estimation errors below $3.5\%$. 
It is also shown that a significant amount of power can be saved by decreasing the input video resolution, which may, however, negatively affect the user experience.


Future studies may attempt to replicate and test our proposed methods and settings on other devices and software applications including wired professional VR headsets like the Oculus Rift or the HTC Vive. Interactive applications including user feedback can also be modeled and analyzed. Finally, the resulting models and power measurements are helpful tools to develop energy and power efficient VR solutions.

\section*{Acknowledgment}
This work was supported by Mitacs and Summit Tech Multimedia ({https://www.summit-tech.ca/}) through the Mitacs Accelerate Program.

\bibliographystyle{IEEEtran}
\bibliography{literatureNeu}

\end{document}